\documentclass[a4,notitlepage,twocolumn]{revtex4-2}
\usepackage{graphicx}
\usepackage{epstopdf}
\usepackage{longtable}

\begin{document}
\title{Is quantum mechanics creationism, and not science?}
\author{Werner Hofer}
\email{werner@wernerhofer.at}
\affiliation{Institute of Physics, Chinese Academy of Sciences, Beijing, China}
\begin{abstract}
I revisit the Bohr-Einstein controversy of 1935. Bohr's assertion that there are no causes in atomic scale systems is, as a closer analysis reveals, not in line with the Copenhagen interpretation since it would contain a statement about reality. What Bohr should have written is that there are no causes in mathematics, which is universally acknowledged. The law of causality requires physical effects to be due to physical causes. For this reason any theoretical model which replaces physical causes by mathematical objects is creationism, that is, it creates physical objects out of mathematical elements. I show that this is the case for most of quantum mechanics.
\end{abstract}

\pacs{PACS numbers:}

\maketitle

\section{A different kind of research}

This article, which covers the last two years of my research, did not result from the usual
working practices of a theoretical physicist. Such practices typically involve the scribbling
of mathematical symbols on a whiteboard or a piece of paper, the rearranging and replacing of these
symbols and their evolution until a final set of symbols is reached which seems to make sense
and can then be put between the text lines in a scientific article.

This research evolved more or less out of a deep meditation on a few words in Bohr's original paper from 1935,
where he tried to prove Einstein wrong, who had accused quantum mechanics of being incomplete \cite{ein1,bohr1}. The seven words in question are 'renunciation of the classical ideal of causality', and it took me quite some time to make sense of them. Renunciation to me sounded ominous, and it is perhaps no coincidence that an emissary of the Pope is a Nuntius, which could explain my slight trepidation, since I was raised a Catholic. The other two words which took a long time to settle in my thinking were 'classical' and 'ideal'. It was not immediately clear to me what these two words had to do with causality. In fact, the more I meditated on them, the less sense they made. In my understanding, which had been philosophically trained on Kant, Schopenhauer and Nietzsche, causality was due to the fact that events happened in space and time and that one event could cause a subsequent one. What this had to do with 'classical',
which I understand to mean, in music, the period from about Haendel to Beethoven, was not obvious. If the 'classical'
posed problems, it became even worse for the 'ideal', because since when were causes conditioned by something 'ideal'? Things happen, because something causes them, which seems to be pretty much accepted across all disciplines in science and engineering. Why this should pose a problem for atomic physics, remained a mystery.

That is, it remained a mystery, until I began looking for causes in atomic physics and did not find any. The theory
describing it only took you so far with causes, and at this point it all became mathematics. Call me suspicious,
but this is exactly what a magician would do: he would lure you into a comfortable feeling telling you how he progresses with his magic trick, until at some point the trick is done and you don't know how it happened. This was,
more or less, the case in all events in atomic physics I analysed. But as a consequence I did not, like most of my
colleagues did over the years, sigh and get on with my mathematical calculations \cite{mermin1}, but started to ask:
Why are there no causes in atomic physics? and What causes physical effects in atomic physics? It took me quite some time to find simple answers to these two questions.

The answers, and the whole story how physics came to lose its causes are written down in a popular science book, which is about to be published \cite{hofer1}. However, I thought that my colleagues, and those too busy to read a reasonably priced book of 300 pages, might appreciate a free and much shorter executive summary. This is what the rest of the article delivers.

\section{Mathematics and other languages}

There are two famous statements about the relationship of mathematics and reality. The first is from Galileo Galilei, when he said that "Nature is written in the language of mathematics"\cite{galilei}. His statement is based on the acceleration of mass in gravity fields and the observation that acceleration is constant and that the path a mass covers is proportional to the time interval squared. This, of course, is described very accurately in the laws of classical mechanics, developed by Newton after Galilei's death.
One can interpret the statement in two possible ways. The first would be that mathematics is a language like any other. The difference being that it is not organised along the laws of a particular grammar, which govern languages, but according to the laws of logic, which govern mathematics. In principle, however, there is no difference between mathematics or any other language.
The other interpretation would be that mathematics has a special relationship with Nature. This interpretation, as will be seen, has led to some confusion in the scientific community, as it often is taken to mean that mathematics somehow provides a closer match with reality than other languages could.

Fortunately for scientists in the 21st century, not even mathematicians claim that their mathematical objects are physical objects in real space, possibly with one exception, see below. If any mathematician would make this claim, then she would have to explain why,
for example, nobody has ever seen a perfect triangle float in space. So it is fairly safe to assume that
mathematics is a particular form of language.

This has been the topic of a heated debate in the middle ages which has, at the beginning of the modern age, led to a clear solution. The debate in the middle ages was between factions which were then called the 'realists'  and the 'nominalists'.
Realists believed that the elements of language were real. So they searched, quite understandably, for the original language, the one that would reflect Nature closest, which they thought might have been the language Adam and Eve spoke in paradise. In their view, elements of this original language are real, and they exist in the real space of our everyday life. Realists traced their belief back to the Greek philosopher Plato.
Nominalists, on the other hand, did not believe that language was real. For them the elements of a language were a social construct and due to a common agreement about their exact meaning. This faction can be traced back to William of Ockham, of Occam's razor fame.

Nominalism is the common principle agreed by modern science in the 17th century. Language, according to this agreement, is a social construct to enable communication between humans and does not exist in reality. Mathematics, according to this understanding, is also not part of reality.

This consensus has been violated by physics in the 20th century. Initially, this was probably clearest expressed by Eugene Wigner when he said that: "The first point is that the enormous usefulness of mathematics in the natural sciences is something bordering on the mysterious and there is no natural explanation for it. Second, it is just this uncanny usefulness of mathematical concepts that raises the question of the uniqueness of our physical theories"\cite{wigner}.
The two points Wigner raised are very profound. Because it is indeed the question why ready-made mathematical concepts like Riemann geometries, Hilbert spaces, Hermitian matrices, Lie algebras and the like would be useful to describe reality in physics.

It will be seen in the following sections that the reason for this feature of physics, in particular quantum mechanics, is that it is largely devoid of causality and ascribes the ability to have physical effects to elements of the mathematical language. The belief that mathematical objects, which do not exist in real space, can cause physical effects in the real world, is very similar to the belief that a God, who does not exist in the real world, created this world. Philosophically, there is no difference between a Bible-belt Christian, who wishes fire and brimstones onto the infidels, and a University quantum theorist, who wishes mathematical symbols to change reality: they are both, at heart, creationists.

Let me finish this section with a quote from MIT physicist Max Tegmark, who seems to have completely lost the ability to differentiate between a language describing Nature and Nature herself. Tegmark's main contribution to the debate is the Mathematical Universe Hypothesis (MUH), which says that "Our external physical reality is a mathematical structure"\cite{tegmark}. Interestingly, his argument is based on the assumption that there exists an external physical reality independent of us humans. The present article is based on exactly the same assumption. Unfortunately, Tegmark's hypothesis that the objects of mathematics are in fact objects in the real world is exactly the same as the assumption of medieval realists. This hypothesis does not agree with the nominalist view, which is the basis of modern science. Linguists will probably also have a word to say about such an idea.

The point to remember from this section is that language is a description of Nature or reality, which in itself has no reality and depends on common consent. Mathematics, it should also be remembered, is only one particular kind of language.

\section{The case against Bohr}

To an unbiased observer, a mechanical engineer, say, or a sociologist, it must seem strange that there exists a distinct difference between experiments and theory in modern physics. While experiments at the atomic scale have improved and the precision and abilities to manipulate systems reached dizzying heights which really seem to allow us to create new materials and new structures from the bottom up, theory seems to be stuck in a time warp, which always reverts back to the 1930s. A historian would probably conclude that something happened in the 1930s, which theoretical physics still has problems to overcome. It is hard to overlook the similarities in the political sphere where, for example, the members of the Orange Order in Northern Ireland still march on the 12th of July every year to reassure themselves by a commemoration of the Battle of the Boyne, which happened in 1690 \cite{boyne}. The reason for this, I would suggest, is the controversy between Einstein and Bohr, and the fact that this controversy transformed theoretical physics from a tool to describe reality by mathematical means to an ideology, which henceforth sought by every means possible to police the opinion that there is no mathematical theory beyond quantum mechanics. Here is, what happened.

We shall come to the case against wavefunctions in the next section, but wavefunctions, by about 1930, became a problem for the logical analysis of events at the atomic scale. Einstein, with his colleagues Podolsky and Rosen, pointed this out in an article in 1935 \cite{ein1}. The following is a slightly simplified description of the measurements and processes Einstein, Podolsky, and Rosen (EPR) considered.
They assumed that two electrons are emitted from an atom in two opposite directions at very high speed. They still retain their common wavefunction as they fly along. Their wavefunction has one component, which is spin-up and one component, which is spin-down. But since these two electrons still have a common wavefunction, we do not know, which electron is spin-up and which electron is spin-down.
To find out, we position magnets in the path of the electrons, at equal distance from the atom, which measure their spin. Magnet A, along the right path at a very large distance from the atom emitting the electrons. And there is magnet B, along the left path also at a very large distance. As magnet A measures the spin of the electron, it is observed that the electron is pushed upwards: this, says a physicist observing the result, is proof that the spin state of the electron arrived at A is spin-up.

But here comes the rub: since the two electrons together have zero spin, the spin of the electron measured at magnet B must be spin-down.
The physicist at magnet A could either measure the spin or not. If she does not, then her colleague measuring at B could obtain both results, either spin-up or spin-down. If she does, then her colleague can only measure the opposite spin to the one measured at A. So clearly, the measurement at magnet A does change the measurement at magnet B.
However, the two measurements are close to simultaneous, as the two electrons have the same speed. So, there is no time for any information from A to reach B in the available interval.

This, said EPR, was proof that either quantum mechanics violated the principle, that nothing is faster than the speed of light, or that there was additional information, which was not contained in the wavefunction and which made the two measurements related.

This problem has now been solved within the model of extended electrons, the solution was recently published \cite{hofer2}. The correct answer is that each electron will carry a phase information, which is revealed at the moment of measurement and which connects the two measurement events. The same is true for photons. So wavefunctions do indeed contain additional information, their phase, which is not considered a physical property in standard quantum mechanics, but nevertheless included in the mathematical description of the problem. The additional information EPR referred to was there all along, but not considered as such.

Bohr, of course, did not know this when he answered that Einstein’s arguments "would hardly seem suited to affect the soundness of quantum-mechanical description, which is based on a coherent mathematical formalism covering automatically any procedure of measurement like that indicated" \cite{bohr1}. And, of course, reality "must be founded on a direct appeal to experiments and measurements" \cite{bohr1}. This statement is the statement of a creationist. It is helpful to first analyse the main components of the statement, and then translate the statement into another language to see the problem.

From the viewpoint of a logical analysis, two terms are problematic in the first sentence. These two terms are "coherent" and "automatic". What they indicate, without any formal proof, is that a mathematical construct exists, quantum mechanics, which is a comprehensive description of reality (this is what coherent implies), and that this construct inevitably leads to all possible measurements one can think of (this is the meaning of the word automatically). Apart from the quite stunning arrogance of the sentence, considering the development of experimental methods since Bohr's statement, which have nothing to do with the experiments which could be undertaken in 1935 (the electron microscope, for example, was only invented in 1936), this statement is not science, but close to religion, as a translation into another language reveals.

Assume that this is not the statement of a physicist, but a novelist, who claims that her book is "a coherent story of the world which covers automatically every possible future, humans can experience." If you ask yourself what sort of book this would be then you reach the conclusion that it probably is not a scientific textbook of any discipline, but rather the Holy Book of one of the great religions, the Bible, or the Quran, for example. So Bohr claimed, to be clear about this point, that a mathematical construct would, for all eternity, be a correct description of all possible aspects of reality in physics. And then he went even further.

"Indeed, the finite interaction between object and measuring agencies conditioned by the very existence of the quantum of action entails - because of the impossibility of controlling the reaction of the object on the measuring instruments ... – the necessity of a final renunciation of the classical ideal of causality and a radical revision of our attitude towards the problem of physical reality" \cite{bohr1}. Let us forget for a moment that causality is and has been a fundamental principle of all sciences for at least three hundred years. Because this is, where Bohr falls foul even of his own Copenhagen interpretation. One of the key statements of this interpretation is captured in the two sentences: "It is wrong to think that the task of physics is to find out how Nature is. Physics concerns what we say about Nature" \cite{bohr2}. But if this is the case then a "renunciation of causality" is not possible, because it would explicitly say that there are no causes in Nature. If this is not possible, then the question remains why there are no causes in quantum mechanics. This will be the topic of the next two sections.

To sum up this section, Bohr's reply to EPR is not only outdated today because the EPR problem has been solved by a causal model in real space. It also contains dodgy logic if compared to his own Copenhagen interpretation, is based on creationism when it assumes that a mathematical construct generates all aspects of reality, and contains assertions which are clearly not science, but religion. I suggest we bury his statements in the history books and get on with the science.

Einstein's view, by contrast, that quantum mechanics is an incomplete theory of atomic physics is vindicated. But the omission goes far beyond what even Einstein thought. It does not concern single elements of reality, which are missing in quantum mechanics, but a whole class of physical objects, which would allow to refer events in real space to physical causes. We shall see, how this works in practice in the next sections, but the final score sheet of the Bohr-Einstein controversy then reads: Einstein one, Bohr nil.

\section{The case against wavefunctions}

One has to be very clear what wavefunctions are, and what they are not. They are not, as Schr\"odinger thought initially, physical objects in real space like electromagnetic fields. This seems sometimes confusing, because the formalism looks very similar to the formalism in Electrodynamics, in particular if the wavefunction is written as a function of location, like $\psi\left({\bf r}\right)$. There is a simple way to distinguish physical objects in space from mathematical objects, and the key question to differentiate is: Does this object contain energy? Every electromagnetic field contains energy, as does mass via the energy-mass relations. The wavefunction, by contrast, does not contain energy. Therefore it is not a physical object in space, but a mathematical object. This is very clear in the abstract formalism, where wavefunctions are objects in their own mathematical space, Hilbert space.

There are two fundamental problems with wavefunctions. The first problem is widely recognized in the community, and it has refused to go away, despite years of hard work by a large number of theoretical physicists. The second problem, which has not been recognized at all so far, is probably the much more important one. These two problems, combined, make wavefunctions not only contradictory entities, but elements of creationism.

The first problem is called the measurement problem and considered by physicists who really think hard about their science one of the fundamental problems in modern physics. The publications and various attempts to solve it are well documented in the literature, and the number of articles trying to account for it probably goes into the thousands. The problem is due to the fact that in a measurement the wavefunction is thought to collapse to its measured state. While this can be stated, it cannot be described consistently in the mathematical framework of quantum mechanics. There is, fundamentally, neither a cause, nor a physical model which would describe how this happens. This problem has led to increasingly weirder speculations about the relationship between the act of measurement, and physical reality, the weirdest one probably the assumption that every measurement creates a new universe. Physicists of this persuasion no longer talk about reality or the universe, but a multiverse which, according to some estimates, contains about $10^{100}$ universes. For an engineer this would probably indicate that physicists have lost their mind and that they are hallucinating weirder and weirder theories to account for a problem that their science seems incapable of solving.

But the problem can actually be turned on its head by two simple questions: What if there is no collapse? How do I measure what I measure without a collapse? My colleague Thomas Pope and I have developed a model of spin measurements based on these two questions, and it turns out that the problem can be solved with two simple assumptions: (1) The electron is an extended object in space, and (2) a magnetic field rotates the spin of electrons, which turns out to be a vector field. The solution has recently been published and presented at various conferences in 2017 \cite{hofer2}. This solves the first problem, and it shows that not the wavefunctions, but the densities are the crucial physical variables leading to a solution.

The second problem, which so far has been completely ignored, is the following little equation, which is due to Max Born \cite{born} (I ignore the various physical units that will usually be added):

\begin{equation}
\psi^{\dagger}\left({\bf r}\right) \psi \left({\bf r}\right) = \rho \left({\bf r}\right)
\end{equation}

Let us be clear about the meaning of this equation. The two objects on the left, the wavefunction and its dual, do not contain energy, they are objects in mathematical space. The object on the right, the density of electrons, contains energy, because electrons have mass, and it is an element not of mathematical space, but of real space. So this equation says that one can take two elements of mathematical space, multiply them, and one creates an object in real space which contains energy. Every time, says the equation, a theoretical physicist takes the square of the wavefunction, energy magically pops out of Hilbert space and appears in real space. This is, without doubt, creationism in its purest form.

Now a traditionalist might try, at this point, to stall the analysis by claiming that the density is not really a physical object, because, after all, it only shows up in statistics. This might have been a valid argument in 1935, but it is no longer relevant in 2018. There are two reasons for this change. The first is that every year thirty thousand scientific papers are published which are based on a theoretical method called density functional theory (DFT) \cite{HK}. In DFT the only physical variable, which determines all physical properties of an atomic scale system is the density of electrons. This firmly roots the density in real space and as a continuous variable. The second reason is that density itself cannot be a statistical property, because this assumption is in conflict with high-precision experiments on metal surfaces, as shown in 2012 \cite{hofer3}. The electron density, not the wavefunction, is the primary physical variable of atomic scale systems. And it is a physical object in real space, not a mathematical object in Hilbert space. So the conclusion remains: the equation describes an impossible relationship of mathematical objects in Hilbert space and physical objects in real space. It is creationism, not science.

The measurement problem, which also is a fundamental obstacle to understanding what happens in atomic scale systems and within the framework of quantum mechanics, can be ignored for most applications of the theory, by following Mermin's recipe to shut up and calculate. This problem, however, cannot be ignored. Because it says, in a nutshell, that quantum mechanics is fundamentally not a causal theory. Not because there are no causes in Nature, which is what Bohr had tried to argue, but because at the point where one commonly expects a cause in a theoretical framework, one gets a mathematical object in Hilbert space.

Within the standard framework, and contrary to the framework of DFT, there is no cause that would make the density attain a particular value. This problem, it turns out, is not only unsolvable, it also makes quantum mechanics unscientific. Creationism is not science, rather the opposite, whether this is within the context of a religion, or within the context of mathematics.

\section{The case against spin}

The spin angular momentum, or spin as I will call it in this section, is probably the most difficult concept introduced in quantum mechanics. The difficulty arises from the fact that it cannot in any way be captured by an image in real space. The usual image, a vector which points up or down, which is used in many scientific papers, does not do it justice, as spin is isotropic in the absence of a measurement, therefore not a vector. The problem, in a nutshell, is the following.

In a Stern-Gerlach experiment silver atoms are detected after an inhomogeneous magnetic field at two distinct points: one point off centre in the direction of higher field strength, one point in the direction of lower field strength\cite{sterngerlach}.
The result indicates that the outermost electron of a silver atom has exactly two possible magnetic dipoles, none of them due to the electron orbit. The original experiments were done with silver atoms, but were later repeated with hydrogen atoms with identical results. What actually happens to the electron in silver or hydrogen has until recently been unknown and it poses quite an interesting logical challenge.
Assume that the electron's magnetic dipole points in a particular direction, and assume that it is random, then the experimental result must be an extended blob. One could now assume, that the direction is not random, but that one class of electrons has a vector which points up, another class has a vector which points down. This would agree with the experiments.

But if the magnet, which determines the trajectory of the atoms, is turned by a quarter rotation, one would measure the exact same result: two points where the atoms impinge on the detection plate, now the points are offset left and right. If the vectors have a particular direction, then every possible direction would make the experimental results different for different rotations of the magnet. This means the vector cannot have a particular direction.
Since it is a fundamental property of every vector that it points into a specific direction, a vector which does not point into a specific direction is a contradiction: so whatever determines the magnetic properties of a hydrogen electron is not a vector, but isotropic. So the electron seems to have a magnetic dipole, which is not only not a vector, but a magnetic dipole which only expresses itself as a vector if it is measured. Both problems have remained profound difficulties for the understanding of the electron until very recently.

The main problem becomes obvious if one asks a simple question: What pushes the silver atom up (down)? The only answer to this question, which is physically possible, is: A magnetic moment, which interacts with the inhomogeneous field of the magnet. But as spin is isotropic, it cannot be a magnetic moment which is a vector. Therefore one has to ask, how an isotropic object becomes a vector, and by what physical process? Described in this way it is obvious that there is no physical process. Instead, there is a similar transformation from mathematical space to real space and from a mathematical object to a physical one. Only in this case it does not involve, as it did for wavefunctions and densities, the creation of energy from Hilbert space, but the creation of a magnetic moment from Hilbert space.
In the Pauli equation the relevant term which accomplishes this creation is the so called Stern-Gerlach term \cite{pauli} (the last term on the right).

\begin{equation}
i \hbar \frac{\partial}{\partial t} | \psi \rangle =
\left( \frac{\left({\bf p} - q{\bf A}\right)^2}{2m} - q \phi\right)I |\psi\rangle - \frac{q \hbar}{2m} {\bf \sigma} {\bf B} |\psi \rangle
\end{equation}

Again, there is no physical cause for the magnetic moment to have a particular direction, there only is a mathematical object with the necessary properties, in this case a Hermitian matrix of a two-dimensional Hilbert space, which accomplishes the transformation. It is, fundamentally, an act of vector-creation from Hilbert space.

One can trace back the logical difficulty to account for these measurements to a single property, tacitly assumed in all of quantum mechanics: the assumption that electrons are point particles \cite{hofer4}. If the electron is a point, then its spin cannot be a vector, because it would have to point into one specific direction. If, however, the electron is an extended object, then spin can be a vector, or rather a vector field. In this case it can also be an isotropic vector field, for example pointing into the radial direction of a sphere. Then the question what pushes the silver atom up or down, is easy to answer: the rotation of the vector field into the direction of the magnetic field vector. Such a process automatically aligns the spin direction with the direction of the external magnetic field, and it will lead to trajectories which are influenced by the field gradient. The model has recently been introduced and it has been shown that it is free of the usual paradoxes \cite{hofer2}.

To sum up the result of this section we find that accounting for spin measurements in quantum mechanics also entails an act of creation, the creation of a vector from a Hermitian matrix in Hilbert space. Again, this is creationism and not science.

\section{The case against Bohm}

On the face of it, the antagonism against Bohm's reformulation of the Schr\"odinger equation in the 1950s seems quite stunning \cite{bohm1}. Because all he really did, was to rewrite the Schr\"odinger equation using a particular form of a wavefunction $\psi = R \exp\left(i S/\hbar\right)$, where both $R$ and $S$ are real-valued variables. This decouples the real and the imaginary components of the Schr\"odinger equation. If one now compares the real part of the equation with the Hamilton-Jacobi equation of classical mechanics, one finds an additional potential, which is commonly called the quantum potential $Q$ \cite{bohm2}:

\begin{equation}
\frac{\partial S}{\partial t} = - \left[\frac{\left|\nabla S\right|^2}{2m} + V + Q \right] \qquad Q = - \frac{\hbar^2}{2m}\frac{\nabla^2 R}{R}
\end{equation}

It has the dimension of a potential like the electrostatic potential $V$. Contrary to a conventional potential like $V$ it does not depend on the physical environment of an electron, but on the shape of its wavefunction via the second derivative of the amplitude $R$. Note that at this point the equations Bohm derived are not different from the original Schr\"odinger equation, because they have been obtained by a general ansatz for the complex valued wavefunction $\psi$, and a simple analysis of the real and imaginary parts of the ensuing equations. The imaginary part can be linked to the continuity equation.

There is quite a large community of physicists who consider themselves Bohmians, and it is indeed tempting to assume that all that makes quantum mechanics different from, say, classical mechanics, is a special potential which only shows up in atomic scale systems. Since this picture is, intuitively, much more satisfying than simply following the agreed recipe and calculating things without being able to picture them in the mind, it is hard to reject out of hand. However, if one accepts that this potential is what makes quantum mechanics different from our everyday environment, then one will have to accept the properties of the potential also as something which belongs to the quantum domain and is not found in an everyday environment. This is, where things become difficult intellectually.

The first problem arises, if one considers the elements which make up the quantum potential $Q$. A potential, which changes the energy content of space where it exists, is always a physical object in real space. This is, why Bohm originally called his theory the "causal" interpretation of quantum mechanics. The point-like electron, he thought, would change its trajectory according to the value of the potential. However, the components making up the potential are the amplitude of the wavefunction and its derivative. The wavefunction is, as already emphasized in previous sections, not an element of real space, but an element of mathematical space, Hilbert space, and it does not contain any energy. So the first problem is again a problem of creationism: the potential is created from elements of mathematical space, energy magically pops out of Hilbert space and into real space. This is not the only problem, though.

Because the relationship between the amplitude of the wavefunction and the quantum potential means that this potential exists throughout the whole space, where the wavefunction exists, and it will change immediately if, for example, the physical environment changes at one point of the system. The wavefunction then will not only change at this point but, via the second derivative and the normalization contained in the quantum potential, it will change throughout the system. Bohm's quantum potential is also non-local. If one now thinks of interactions between electrons via electromagnetic fields, then the electromagnetic fields will only interact with electrons to the extent that they have time to propagate to the point of interaction. The quantum potential, however, will interact with an electron instantly.

So while on the one hand the image that a special potential is what makes quantum mechanical systems different from other physical systems is intellectually satisfying, it is on the other hand hard to accept that in this case one will have to give up causality, because this potential is created from mathematical space and there is no physical mechanism which would allow me to understand how this quantum potential actually operates in space. Bohm's reformulation of the Schr\"odinger equation leads to exactly the same problem as in the original theory: there are no causes.

This could be the endpoint of the analysis and one could then conclude that Bohm's theory is probably not a way to regain causality in quantum mechanics. But one could also go one step further. If it is accepted that Bohm's theory is non-local and a-causal, then one could ask what this means for the original theory described by the Schr\"odinger equation. Formally, Bohm's equations are not different from the original equation, because his ansatz for the wavefunction is generally valid.

It is then hard to see how one set of equations makes a theory manifestly non-local, while the logically equivalent set of equations does not have this deficiency. So the question remains: Is wave mechanics itself non-local? Tentatively, I would suggest that the answer to this question is yes. Then one has to understand where this non-locality would come into the original theory. The best candidate, I think, for this is the fact that one obtains physical properties of electrons, which are considered point particles in the conventional theory, only if the operator equations are integrated over the whole space of the wavefunction. So, for example, for the hydrogen electron this means an integration over infinite space, since the wavefunction exists as an exponentially decaying function over the whole space. I suggest that this procedure, the integration over infinite space, makes wave mechanics as non-local as Bohm's theory, where non-locality is made explicit in the form of the quantum potential. It is understood that non-locality also makes a theory a-causal. If this argument is correct, then what Bohm did was not to invent a new theory which would allow us to better understand atomic scale systems, he rather revealed that there is no way one can make quantum mechanics a framework based on causality.

\section{Predictions and correlations}

If there are no causes in quantum mechanics and if all physical effects are created from mathematical objects, that is elements of the mathematical language, what does this entail for the science described by quantum mechanics?  A brief recourse to history will make it clear.

During Galilei's lifetime the conventional wisdom in astronomy was that the solar planets move around the Earth on trajectories described by epicycles on top of circles. The observations of Mars, for example, would show exactly such a behaviour. Also in this case the physical cause for its motion was unknown. So the mathematical model did not connect physical causes with physical effects, it connected mathematical objects (circles) with physical effects. Logically, what this theory describes is not a set of mathematically formulated predictions how the planet moves, but only a correlation between an observation (the position of the planet) and a mathematical model (circles). One part of the astronomical data could have, even before Kepler's observations, given away the fact that all was not well with epicycles, and this part concerned the velocities of planets along their trajectories, which were not constant. In atomic physics, the change of the wavelength of electrons as they change their velocity is also a fact that remained unexplained in quantum mechanics and could have alerted physicists long ago that all was not well in theoretical physics.

Predictions can only be made, if a mathematical model relates a physical cause to motion or other physical effects, and hence observations. Only Newton's theory of gravitation, which provided these causes forty years after Galilei's death, is capable of making these predictions.

The same applies to quantum mechanics. The mathematical models do not connect physical causes to physical effects, because there are no physical causes. The theory is therefore in principle unable to make predictions. All it provides are correlations between mathematical models and experimental observations.  Historically, it involved the invention of mathematical objects to account for experiments. First, the invention of matrices in Heisenberg's matrix mechanics, then the invention of wavefunctions and spins in wavemechanics. Not one of these objects is a physical object in real space. A similar case in point, to be analysed in the future, will probably be provided by particle physics, which was forced to add a plethora of new mathematical elements, called particles, as the experiments progressed. Given that it is always possible to add new mathematical elements to the description if a model is not in line with observations, there is also no way to falsify such a theory. Quantum mechanics would, logically speaking, also fail Popper's test for a valid scientific theory.

Quantum mechanics, in short, is not science.

\section{Conclusions}

The inevitable conclusion from the analysis in the preceding sections is that a major part of modern physics, quantum mechanics, is creationism, in principle not falsifiable, and not science. This generates an interesting set of problems for theoretical physicists. One way to deal with the problems is to ignore the findings and to try to discredit the author of the paper. This is, what the establishment in physics did rather successfully with David Bohm, and it is probably safe to say that this will be the first reaction.

But will theoretical physicists be able to keep a straight face and the necessary authoritative demeanor when they teach quantum mechanics 101 in the future? Will they be able to stifle a snigger when they write down Born's equation or multiply a Pauli matrix with a field vector to obtain a magnetic moment? Not to mention the awkward possibility that students might start to call their Physics professors colloquially professors of Creationism, and rightly so. If this situation is already quite difficult to handle for a real scientist, the second problem is even worse. Because what will biologists think, who had to fight against creationism ever since Charles Darwin published his book? One can predict that physics, as a science, will lose much of the respect it currently enjoys in the scientific community. This leads to the third problem, which is finding a way to make physics a real science again. Here, the question is how much will have to be changed, and how much of the current conventional wisdom will have to be discarded for a future, strictly scientific, physics.

There is, unfortunately, no easy way out. The whole problem of creationism should have been addressed eighty years ago and not swept under the carpet by the faithful followers of Bohr. It should never have been allowed to fester and to impact on all subsequent theory.

An interesting question, which historians might want to investigate, is how much of the analysis in the current article had been understood by Bohr himself. And if he understood the consequences, how did he think that a lack of causes could lead to a theory which describes physical effects in a scientific manner? How did he think he could keep divine mathematical interventions at bay which, as shown here, permeate the very foundations of the theoretical framework? My suspicion is that, in the end, it might come down to nothing more than personal arrogance, a lack of scientific humility, and a conviction to always be right.

For me the most frightening aspect of this analysis is what it says about us physicists. If quantum mechanics, which is one of the corner stones of modern physics, is actually not science but creationism, then how can we justify teaching our students the same nonsense? What they signed up to, when they entered University, was to get an education in a science discipline which gives them the expertise to understand and to work with the reality they are living in. Teaching them creationism, and calling it science, is irresponsible. So I would urge all of my colleagues in theoretical physics to analyse their own field along the same lines. Not by mindlessly heaping mathematical symbols onto a whiteboard and then, at some point, magically finding physical objects, but by analysing  whether what their theory does is actually compatible with the laws of causality. If it is not, it has no place in science. My feeling is that this will probably apply to most of modern physics, not just quantum mechanics. Time, I would think, for a big bonfire of theoretical tradition.


\begin{references}
\bibitem{boyne}
A colleague from Northern Ireland has made me aware that the Orange Order actually marches on the 12th of July every year, even though the battle happened, historically, on the 1st of July. The reason for the discrepancy is a change from the old Julian calendar to the Gregorian calendar in the 18th century. Interestingly, Guy Fawkes night, which commemorates an event on the 5th of November according to the Julian calendar, is still celebrated on the 5th of November, possibly because the 17th of November would not rhyme that well.
\end{references}
\end{document}